\documentclass[
reprint,
showpacs,
showkeys,
amsmath,amssymb,
aps,
]{revtex4-1}

\usepackage{graphicx} 
\usepackage{dcolumn} 
\usepackage{bm} 
\usepackage{hyperref} 

\begin{document}

\preprint{APS/123-QED}

\title{Complex self-sustained oscillation patterns in modular excitable networks}

\author{Jason Danison}
\email{jdanison@bmcc.cuny.edu}
\altaffiliation{Previously known as Bogdan Danila.}
\author{Miguel Perez}
\affiliation{BMCC, The City University of New York, 199 Chambers St, New York, New York 10007-1047}

\date{\today}

\begin{abstract}
We study the relationship between the modularity of scale-free excitable networks and their ability to support self-sustained oscillation patterns. We find that the probability for a network of given degree-distribution exponent to be able to support self-sustained oscillations is strongly affected by its modularity. In addition, both high- and low-modularity networks are more likely to exhibit long-period oscillation patterns than those with intermediate modularity, but the degrees of complexity and correlation in these two cases are different. The long-period oscillations cannot be explained by a minimum-length Winfree loop, but instead arise from the interplay between two or more propagating waves. Finally, we introduce a new method that can be used to analyze the structure of the self-sustained oscillation sources at different levels of detail and show that the period of the oscillation pattern is statistically correlated with the fraction of modules that are part of the oscillation source.
\end{abstract}


\maketitle


\section{Introduction}

The propagation of waves in continuous excitable media\cite{Kapral, Meron, ChenEfield, BarEiswirth, Steinbock, Aranson} as well as in discrete lattice structures\cite{QianFundStruct} has been an important research topic for many years\cite{CrossHohReview}. The propagation of spiral and target waves in both of these situations is well understood. More recently, a number of authors have studied the sustained activity patterns in various complex network models\cite{MullerStruct, KanakovClustSynch, MaWave, LiaoPatternForm, KakmeniHindmarshRose, LiLiHu}, including simple small-world\cite{RieckeMultipleAttractors, QianRegulation, SinhaEmergenceSW, QianStructureControl, KobayashiSustained, RoxinSelfSustained} and Erd\H{o}s-R\'{e}nyi networks\cite{QianER, QianTimeDelay, QianMinWinfree}.

The interest in these network models is driven primarily by neuroscience\cite{AvenaCommunDynamics, ZhengDPAD, BazhenovBrain, RulkovBrain, BuzsakiBrain, UsreyVisual, WardCognitive, SteriadeSleepArousal}. In particular, it has been shown that wave propagation in complex neuronal networks is involved in many brain functions such as visual perception\cite{UsreyVisual}, cognitive processes\cite{WardCognitive, AvenaCommunDynamics} and sleep-arousal patterns\cite{SteriadeSleepArousal}. In addition, the presence of high-degree neurons is crucial for the ability of the cortex to perform its information-processing functions\cite{TimmeCorticalComput}. The authors of Ref.~\cite{MiLongPeriodPNAS} describe long-period rhythmic synchronous firing in Barab\'asi-Albert-like scale-free networks and propose a Hebbian learning mechanism leading to topologically similar neuronal networks as the basis for the memorization of information encoded in long temporal intervals. Within the framework of their model, the self-sustained source of the oscillations on the network is assumed to be a single, simple, loop (Winfree loop). A similar situation arises in the case of Erd\H{o}s-R\'{e}nyi excitable networks, where Ref.~\cite{QianMinWinfree} finds that the self-sustained source of the oscillations is a shortest-path loop of a certain minimum length, which depends on the speed of propagation of the excitations. An important question that arises is whether there may be other, more complex, mechanisms for the generation of long-period oscillations, which may not be conditioned by the presence of a long simple loop and which could be used for the memorization of complex temporal patterns. This question is even more justified when one considers the limited range of network topologies that have been investigated until now.

Most studies of excitable networks published so far are based on simple network models which do not allow for independent control over various aspects of network topology like clustering, node degree distribution, degree-degree correlations or modularity. For example, small-world networks are modeled in Refs.~\cite{RieckeMultipleAttractors, QianRegulation, SinhaEmergenceSW, QianStructureControl, KobayashiSustained, RoxinSelfSustained} by randomly rewiring a small fraction of the connections of a two-dimensional nearest-neighbor lattice, which produces neither a scale-free distribution of the node degrees nor a modular structure. The networks studied in Ref.~\cite{MiLongPeriodPNAS} are scale-free but cannot exhibit a modular structure either. In addition, no attempts have been made so far to study the statistics of ensembles of networks with excitable nodes beyond sampling the space of the initial conditions. To address these limitations, we use the network model described in Ref.~\cite{Benchmarks}, which produces ensembles of modular networks with a large array of tunable parameters, of which we focus on network size, average degree and modularity.

Modularity is generally known to have a significant bearing on a network's ability to perform its functions as it has been observed that subsets of a network whose nodes are more densely connected than in a random ``null model" are likely to perform some function together\cite{FortuRev, NewmanDivisive, NewmanPREspect, NewmanPNAS}. In the case of small-world networks of excitable nodes, the main obstacle to the establishment of a self-sustained pattern of oscillations is the rapid spreading of excitation, which places the bulk of the network in a refractory state. Intuitively, it is to be expected that a modular structure will be able to mitigate this effect by limiting the propagation of the excitations, and that it might even lead to situations where wave patterns that are only weakly coupled propagate across separate sets of communities.

It is important to note that small-scale studies have already suggested a correlation between network dynamics and modularity. In Ref.~\cite{LiLiHu}, the authors show that the community structure can be inferred by measuring the degree of synchronization between the nodes. However, they do this only for the case of a network with 128 nodes and 4 equal communities and for three small social networks. Ref.~\cite{MullerStruct} uses a simple SIR model for network dynamics to study the relationship between node centrality or network modularity and the patterns of oscillatory activity, but their results are again confined to a small number of networks.

Finally, the role played by the strength of the coupling between the neurons has not yet been systematically studied, even though it has beed demonstrated that this parameter can affect the behavior of the network\cite{MiLongPeriodPNAS}. In contrast, here we present results for a wide range of values of the coupling strength.

The paper is organized as follows: in Section II we give a brief overview of the concept of network modularity and we describe the dynamic model. This is followed by the introduction of an improved method for identifying the self-sustained source of the oscillations. In Section III we present results detailing the workings of networks of both low and high modularity while Section IV is devoted to results for statistical ensembles of networks.

\section{Method}

\subsection{Network structure and dynamics models}

The concept of community structure arises from the fact that many networks can be naturally divided into subsets of nodes such that the density of connections within these subsets is higher than between them. Networks that exhibit a clear structure of this kind are called modular. The simplest and most straightforward way to quantify the modularity of an undirected unipartite network is by means of the modularity function $Q$ introduced by Newman and Girvan in Refs.~\cite{NewmanGreedy, NewmanPREspect, NewmanPNAS}. This function is defined as
\begin{equation}
Q = \sum_{k=1}^K \sum_{i,j\in C_k} \left( A_{ij}-\frac{d_i d_j}{2m} \right) ,
\label{eq:one}
\end{equation}

\noindent where $A$ is the adjacency matrix of the network, $\{C_k\}$ with $k=\overline{1,K}$ is the set of communities, $d_i$ is the degree of node $i$ and $2m=\sum_{i=1}^N d_i$. If the community structre of a network is unknown, the maximization of the modularity function provides a way to identify it\cite{NewmanPNAS}.

We considered ensembles of random scale-free networks with tunable modularity generated using the algorithm described in Ref.~\cite{Benchmarks}. These undirected networks have a built-in community structure that is provided on output. The properties of the ensemble are controlled by a number of parameters which include the network size $N$, the average degree $\left<d\right>$, the maximum degree $d_{max}$ and the mixing parameter $\mu$, which represents the average fraction of links running between different modules. The other parameters were kept at their default values, including the exponent of the power-law degree distribution $\gamma=-2$. It is important to note that the networks that were not fully connected were rejected.

Network dynamics was defined by a variant of the B\"{a}r-Eiswirth model\cite{BarEiswirth},
\begin{eqnarray}
\frac{du_i}{dt} &=& \frac{1}{\varepsilon} u_i(1-u_i) \left(u_i-\frac{v_i+b}{a}\right) + c \sum_{j\in\mathcal{N}_i} (u_j-u_i) \\
\frac{dv_i}{dt} &=& f(u_i)-v_i ,
\label{eq:two}
\end{eqnarray}

\noindent where $u_i$ and $v_i$ are analogous to the concentrations of activator and inhibitor or to the membrane potential and recovery current, $c$ is the coupling strength between neighboring nodes, $\mathcal{N}_i$ is the set of neighbors of node $i$ and the function $f(u)$ is defined by
\begin{equation}
f(u)=\begin{cases} 0\:\mbox{for}\:u<\frac{1}{3} \\ 1-6.75u(1-u)^2\:\mbox{for}\:\frac{1}{3}\le u \le 1 \\ 1\:\mbox{for}\:u > 1 \end{cases}.
\label{eq:three}
\end{equation}

The system of differential equations was integrated using a fourth-order Runge-Kutta routine with integration step $h=1/128$. Following previous work\cite{BarEiswirth, MiLongPeriodPNAS, QianER, QianRegulation, QianStructureControl, Meron}, we set $\varepsilon=0.04$, $a=0.84$ and $b=0.07$ but we explored a wide range of values for the coupling strength $c$ between 0.1 and 0.7.

\begin{figure}
\scalebox{0.33}[0.33]{\includegraphics*{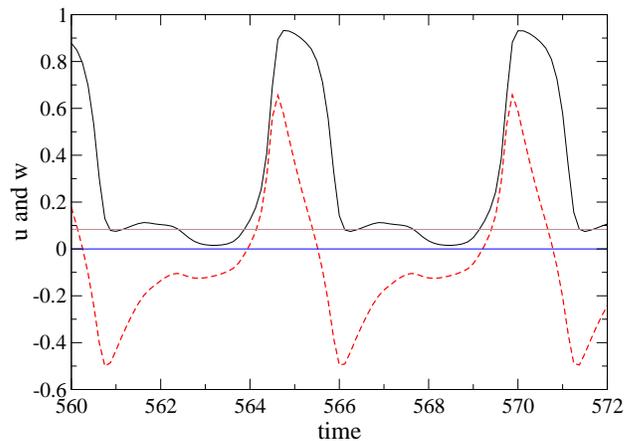}}
\caption{\label{fig:CORRECT64} (Color online) Amplitude $u(t)$  (black continuous) and the factor $w(t)$ (red dashed) for a node where DPAD fails. The horizontal gray line above zero represents the threshold $b/a$.}
\end{figure}

\begin{figure}
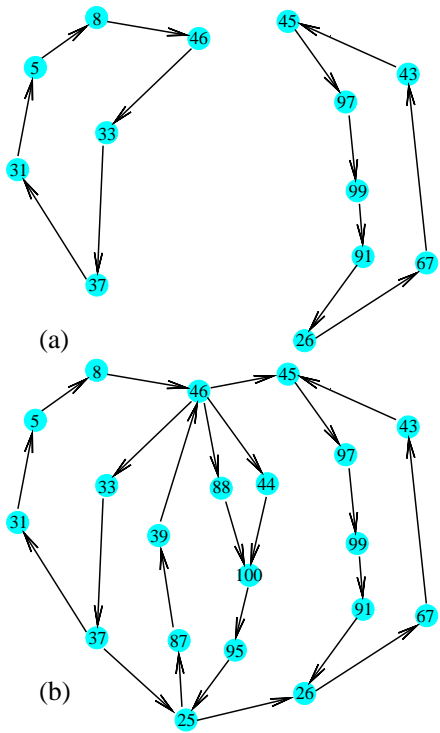

\scalebox{0.4}[0.4]{\includegraphics*{Net-mu30-c50-Core}}
\linebreak
\scalebox{0.4}[0.4]{\includegraphics*{Net-mu30-c50-Loopy}}
\caption{\label{fig:LoopyCore} Different degrees of simplification of the oscillations source for a network with $N=100$ nodes and $\left<d\right>=4$: (a) with maximum individual thresholds $D_{th,i}=D_{i,max}$ and (b) with maximum collective threshold $D_{th}=\min\{D_{i,max}\}$.}
\end{figure}

\subsection{Identification of the oscillations source}

A number of previous studies have used the dominant phase-advanced driving (DPAD) method\cite{QianStructureControl} to identify the source of the sustained oscillations on the network. This method attempts to find the source of the oscillations by retaining only the links between each node $i$ and the neighbor which provides the strongest driving at the moment when $u_i(t)$ crosses the threshold value $b/a$ while increasing. Following extensive testing, we identified many situations where this is not the best choice, since it can lead to a false identification of the dominant driving node. Consider the example in Fig.~\ref{fig:CORRECT64}, which shows the behavior of a node from a network generated with parameters $N=100$, $\left<d\right>=4$, $d_{max}=15$ and $\mu=0.3$ for which the coupling strength was set to $c=0.5$. This is a network with a relatively high modularity $Q=0.565$. A few nodes exhibit two above-threshold maxima of $u$ in the course of a period (continuous line), but only one of these maxima represents true firing. The lower maximum occurs while the node is strongly driven by a different set of neighbors, towards the end of its refractory period, but the concentration of inhibitor $v$ is still too high and $u$ drops as soon as the driving subsides.

A simple way to avoid this problem is to define the dominant phase-advanced node of $i$ as its strongest driver at the moment when the quantity $w_i = u_i - \frac{v_i+b}{a}$ (dashed line in Fig.~\ref{fig:CORRECT64}) crosses 0 while increasing. This provides essentially the same timing as the original method in the case of true firing but avoids the futile firing attempts during the refractory period.

\begin{figure*}
\scalebox{0.5}[0.5]{\includegraphics*{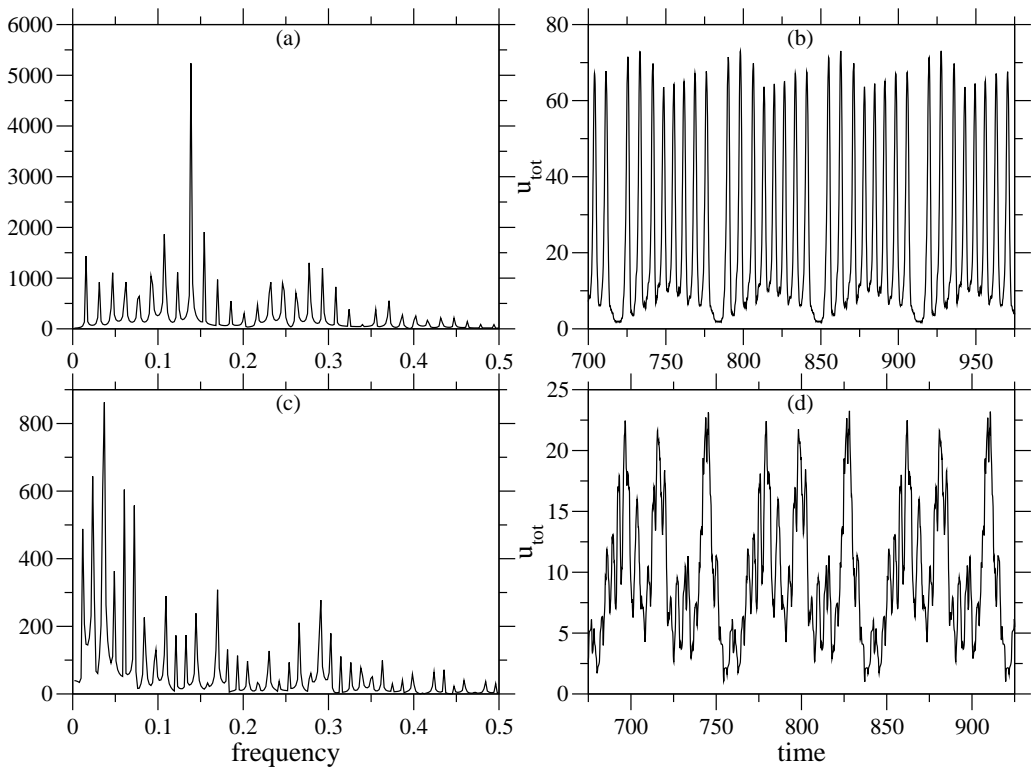}}
\caption{\label{fig:FFT} Discrete Fourier transforms ((a) and (c)) and plots of $u_{tot}$ vs.~$t$ ((b) and (d)) for long-period oscillation patterns of two networks with $N=100$ and $\left<d\right>=4$ but different modularities, $Q=0.404$ ((a) and (b)) and $Q=0.766$ ((c) and (d)).}
\end{figure*}

In addition, there are situations when two or more nodes provide roughly equal driving to a common neighbor and, moreover, that neighbor would not be able to fire at the right time to continue the propagation of the wave without all major contributions. Therefore, it is important to be able to generate subnetworks exhibiting a lesser, variable, degree of simplification. Such a subnetwork would include the links between every node and all its major drivers. One way to achieve this is to build a ``driving matrix" whose elements $D_{ij}$ are the averages of the differences $\delta_{ij}=u_j-u_i$ recorded at times when $w_i(t)=0$ while increasing and then setting all elements below a certain threshold equal to zero. In addition, as a way to focus on the nodes that are essential to the propagation of the self-sustained oscillations, one can remove all the ``dead-end" nodes which never contribute to the driving of any neighbor. This must be done recursively, since removing a set of non-driving nodes may put other nodes in this category.

The detailed procedure is as follows:

\begin{enumerate} {

\item For every node $i$ and every $j\in\mathcal{N}_i$, record the values of $\delta_{ij}=u_j-u_i$ at every moment when $w_i(t)=0$ and $\frac{dw_i}{dt}>0$. If the network's phase space trajectory settles on an attractor, one may wish to consider only values recorded after this has happened.

\item At the end of the simulation, compute the driving coefficients $D_{ij}=\left<\delta_{ij}\right>$. If $i$ and $j$ are not connected set $D_{ij}=0$.

\item For every node $i$, identify the largest driving coefficient $D_{i,max}$.

\item Choose a threshold value $0\le D_{th}\le \min\{D_{i,max}\}$ and if $D_{ij}<D_{th}$ set $D_{ij}=0$. Alternatively, one can use separate thresholds $0\le D_{th,i}\le D_{i,max}$ for every node.

\item For every node $i$ that does not contribute to the driving of a neighbor set $D_{ij}=0$ for all $j$. Repeat this step until no such nodes are found.

\item Symmetrize the matrix D by setting $D_{ij}=\max(D_{ij},D_{ji})$ and, if desired, set all non-zero coefficients equal to 1.

\item Treat the rows and columns of matrix D that contain non-zero elements as the adjacency matrix of the oscillation source subnetwork.

} \end{enumerate}

Two subnetworks representing different degrees of simplification of the network from which the example in Fig.~\ref{fig:CORRECT64} was taken are shown in Figs.~\ref{fig:LoopyCore} (a) and (b), corresponding to setting individual $D_{th,i}=D_{i,max}$ and respectively a global $D_{th}=\min\{D_{i,max}\}$. The arrows in these figures show the direction of propagation of the wave, derived from the coefficients of the ``driving matrix" $D_{ij}$ before symmetrization (Step 6). In Refs.~\cite{QianStructureControl, QianER} the authors describe only situations where the activity on the entire network can be traced to waves propagating along a single Winfree loop\cite{QianStructureControl, WinfreeSurveyBehav}. We see from Fig.~\ref{fig:LoopyCore} (a) that this description is incomplete, since multiple loops may be present within a certain wave pattern. Note that the loop on the right is longer than the one on the left, which means that the network must include a mechanism for their synchronization. Fig.~\ref{fig:LoopyCore} (b) shows a still simple but much more complete picture of the network's workings. In addition to the two main loops, there is a third loop $\{46, 44/88, 100, 95, 25, 87, 39\}$, of the same length as the loop on the right but exhibiting weaker driving. The propagation along the central loop is sped up by early activation of node 25 by node 37. At the same time, these loops contribute to the driving of loop $\{45, 97, 99, 91, 26, 67, 43\}$ through the links $\{46, 45\}$ and $\{25, 26\}$. It is important to note that two of the inter-loop links shown in Fig.~\ref{fig:LoopyCore} (b) are critical for the persistance of the self-sustained oscillation on the network. If one of links $\{46, 45\}$ or $\{37, 25\}$ is removed, no self-sustained pattern can be established. On the other hand, removing only $\{25, 26\}$ still allows a slightly different oscillation pattern.

\section{Results for example networks}

The analysis method described above was applied together with discrete Fourier analysis of the oscillation pattern to study the dynamics of networks with different modularities $Q$ and different coupling strength parameters $c$. We found that many networks exhibit complex oscillation patterns whose long periods cannot be accounted for by propagation along the maximum shortest-path loop, let alone by propagation along the minimum shortest-path loop with propagation time longer than the refractory period (minimum Winfree loop). This picture is very different from the one found by Refs.~\cite{QianMinWinfree, MiLongPeriodPNAS} in the case of Erd\H{o}s-R\'enyi and simple small-world networks. Surprisingly, long-period oscillation patterns are more likely in the case of networks with low or high modularity than in the case of those with intermediate modularity. We also found oscillation patterns that exhbit reversals of the direction of propagation along some of the loops that are part of the oscillation source. It is also important to note that a given network may exhibit a number of stable oscillatory patterns with different levels of complexity.

Qualitatively different types of long-period behavior have been observed in the case of low- and high-modularity networks. Representative results are shown in Fig.~\ref{fig:FFT}. The Fourier transform of $u_{tot}=\sum_{i=1}^N u_i$ and a plot of $u_{tot}$ versus time are shown in Figs.~\ref{fig:FFT} (a) and respectively (b) for a scale-free network of $N=100$ nodes, average degree $\left<d\right>=4$ and relatively low modularity, $Q=0.404$. The coupling strength in this case is $c=0.50$. The oscillatory pattern consists of a series of nonidentical bursts of synchronous firing interrupted by longer periods of low activity, with a period $P=64.8$. This pattern is qualitatively typical for the low-modularity networks that exhibit long periodicity. Note that the same network also supports an oscillatory pattern with a much shorter period, $P_S=7.42$, slightly different from the 7.2 period of the prominent ninth harmonic of the long-period variant. The picture is quite different in the case of high-modularity networks. Results for one such network, also with $N=100$ and $\left<d\right>=4$ but a much higher modularity $Q=0.766$, are shown in Figs.~\ref{fig:FFT} (c) and (d). A lower value of the coupling strength $c=0.15$ was used in this case. The oscillatory pattern now consists of three distinct, less synchronous, bursts, each confined to a different part of the network. This shows that a modular structure may indeed prevent global synchronous firing, instead causing the excitation to cycle through the set of communities. The resulting period in this case is $P=82.6$. The same network also exhibits two periodic oscillation patterns of periods 32.7 and 34.3, as well as sustained non-periodic oscillations.

\begin{figure}
\scalebox{0.33}[0.33]{\includegraphics*{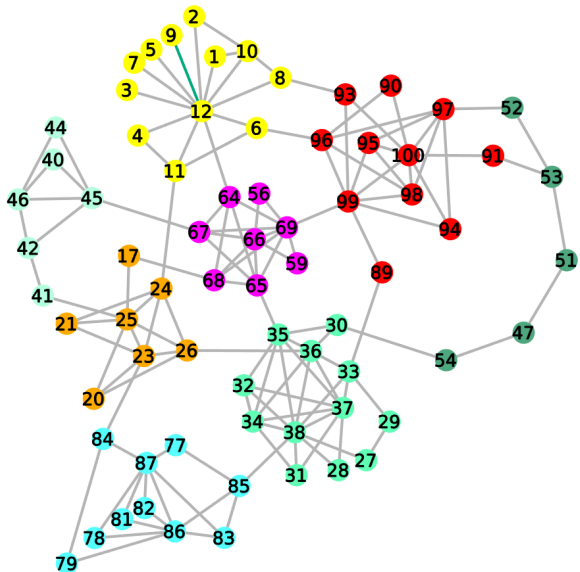}}
\caption{\label{fig:Example-mu10-c15} (Color online) Simplified picture of a network of size $N=100$, average degree $\left<d\right>=4$ and modularity $Q=0.777$ which exhibits a very long period oscillation pattern. The different colors represent different communities.}
\end{figure}

\begin{figure}
\scalebox{0.32}[0.32]{\includegraphics*{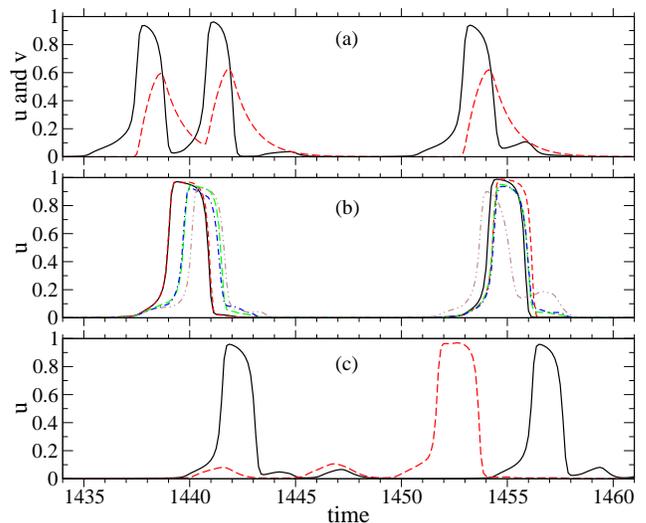}}
\caption{\label{fig:Explain-mu10-c15} (Color online) The backfiring mechanism of the community of node 23 on the network in Fig.~\ref{fig:Example-mu10-c15}. (a) Concentration of activator $u$ (black continuous) and inhibitor $v$ (red dash) for node 23. (b) Concentration of activator for nodes 20 (black continuous), 21 (red dash), 24 (green long dash), 25 (blue dash-dot) and 26 (brown dash-dot-dot). (c) Concentration of activator for nodes 11 (black continuous) and 36 (red dash).}
\end{figure}

Fig.~\ref{fig:Example-mu10-c15} shows a simplified picture of another network with a particularly long period $P=256$. This network has $\left<d\right>=4$, a modularity $Q=0.777$ and the simulation was run with $c=0.15$. For this value of $c$, nodes with a degree of 7 or higher cannot be excited by a single neighbor, while in the case of nodes of degree 5 or 6 successful excitation by a single neighbor depends on the duration of that neighbor's pulse. The simplified network was obtained by setting $D_{th}=0$ to provide the most inclusive definition possible for the oscillation source. The network has 9 communities, out of which 8 are represented in the oscillation source and are labeled by different colors. The workings of the network over slightly more than one period are shown in \cite{Video}. The first thing one notices while watching the simulation is that there is not a single one-way loop. The only loop that ever closes is the one containing nodes $\{51, 47, 54\}$ and the communities of nodes 37 and 100, when travelled counter-clockwise. Clockwise propagation along this loop is blocked because node 99 has too high a degree to be excited by node 89. Nevertheless, propagation in this direction is sometime initiated from the community of node 100, which has a different timing pattern compared to the community of node 37.

\begin{figure*}
\scalebox{0.5}[0.5]{\includegraphics*{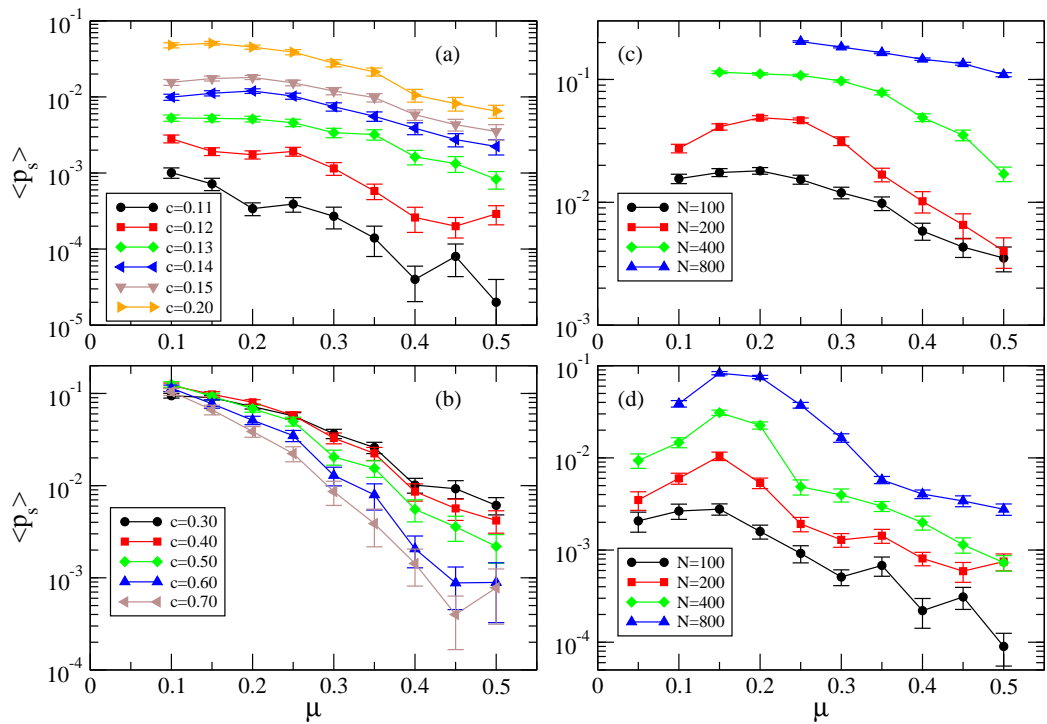}}
\caption{\label{fig:pavg} (Color online) The average probability to find a phase space realization that exhibits self-sustained oscillations $\left<p_s\right>$ as a function of the mixing parameter $\mu$. The results are for $N=100$, $\left<d\right>=4$, $d_{max}=16$ and different values of the coupling strength ((a) and (b)), for $\left<d\right>=4$, $d_{max}=16$, $c=0.15$ and different network sizes (c), and for $\left<d\right>=6$, $d_{max}=24$, $c=0.15$ and different network sizes (d).}
\end{figure*}

Another interesting feature exhibited by this network is that propagation from node 84 to the community of node 23 sometimes ``backfires", depending on the inputs that this community receives from nodes 11 or 36. The mechanism is explained by reference to Fig.~\ref{fig:Explain-mu10-c15}. Fig.~\ref{fig:Explain-mu10-c15} (a) shows the concentrations of activator $u_{23}$ and of inhibitor $v_{23}$ as functions of time for two consecutive pulses of node 23 driven by node 84. The first pulse is quickly followed by a second one, driven collectively by the community neighbors of node 23 (Fig.~\ref{fig:Explain-mu10-c15} (b)), and the wave propagates back to node 84 and its community. Note that the lowest degree neighbors 20 and 21 fire during the refractory period of node 23, but as the wave propagates to nodes 24, 25 and finally 26 the prolonged driving is just enough to cause the second pulse. On the other hand, the next time node 23 is driven by 84, node 26 is driven earlier by the firing of node 36 (Fig.~\ref{fig:Explain-mu10-c15} (c)), which causes the rest of the community to fire too early to drive node 23 a second time. The pulses of node 11 arrive too late in these two cases, but there are other cases when the backfiring is inhibited by this node.

This example proves that the mechanisms involved in the generation of self-sustained oscillations on complex networks can be much more complicated than what was previously assumed and that long periods can arise from the interplay between different propagation times over different parts of the network.

\section{Results for statistical ensembles}

In this section we present results concerning the relationship between modularity and the likelihood for a network to exhibit self-sustained oscillation patterns as well as between modularity and the period of the wave pattern. The statistical ensemble for each set of parameters $N$, $\left<d\right>$, $d_{max}$ and $\mu$ consisted of 100 networks. Each network was started 1000 times with random initial conditions, the sets $\{u_i\}$ and $\{v_i\}$ being independently and uniformly distributed between 0 and 1.

Fig.~\ref{fig:pavg} shows the average probability to find a phase space realization that exhibits self-sustained oscillation patterns $\left<p_s\right>$ as a function of the mixing parameter $\mu$. The average is performed over the 100 realizations of the network ensemble, with the error bars representing the standard error of the mean. Note that a high value of $\mu$ means a low average modularity of the network ensemble, with $\left<Q\right>$ decreasing from about 0.8 to about 0.4 from left to right. The results in Figs.~\ref{fig:pavg} (a) and (b) are for $N=100$ and different values of the coupling strength $c$ ranging from 0.11 to 0.7 while Figs.~\ref{fig:pavg} (c) and  (d) display results for $c=0.15$ and different network sizes $N$ ranging from 100 to 800. The ratio $d_{max}/\left<d\right>$ is the same for all curves but the ratios $N/\left<d\right>$ vary significantly in the figures on the right.

\begin{figure*}[t]
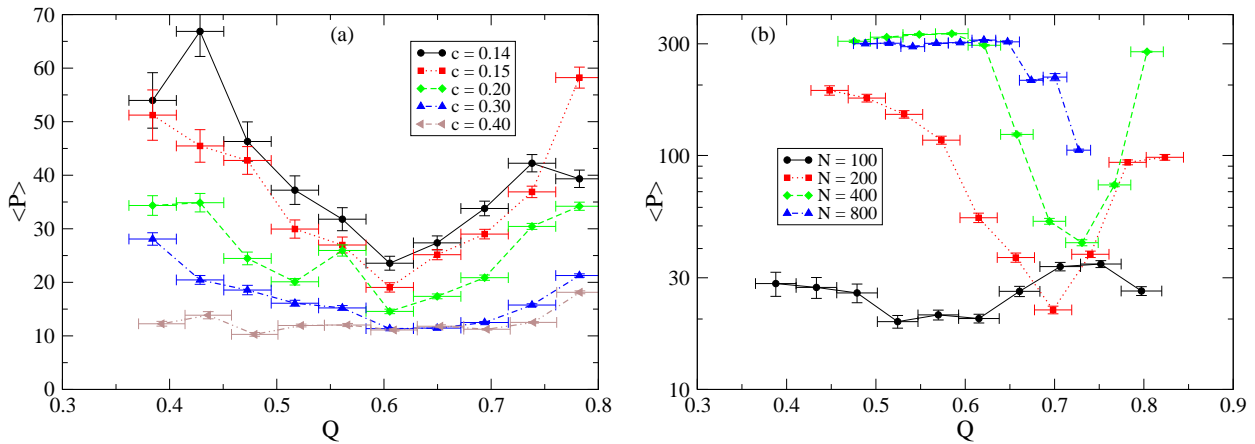

\scalebox{0.33}[0.33]{\includegraphics*{Perl-vs-Q-be3b-N100-K400-mK15-varc-D1k-s-177290new}}
\quad
\scalebox{0.33}[0.33]{\includegraphics*{Perl-vs-Q-varN-K400-c15-D2k-s-177290}}
\caption{\label{fig:Perl-vs-Q} (Color online) Average period in different modularity ranges for (a) $N=100$ and different values of the coupling strength $c$ and (b) $c=0.15$ and different network sizes $N$. The horizontal bars show the extent of the modularity ranges while the vertical bars represent the standard error of the mean. All results are for $\left<d\right>=4$ while $d_{max}$ is 15 (a) and 16 (b).}
\end{figure*}

For $N=100$ the probability for self-sustained oscillations $\left<p_s\right>$ exhibits a rapid overall increase with $c$ between 0.1 and 0.15. Above $c=0.2$, $\left<p_s\right>$ begins to decrease with increasing $c$ in the case of low-modularity networks due to the fact that the hubs can now be excited by the simultaneous firing of a smaller number of neighbors, which is more likely to lead to widespread simultaneous firing followed by exctinction. However, a highly modular structure is able to mitigate this effect and $\left<p_s\right>$ remains high for such networks until it finally starts to decrease above $c=0.7$. The probability $\left<p_s\right>$ increases with network size for given $\left<d\right>$ and $d_{max}$. Note that the maximum value of $\left<p_s\right>$ increases faster with $N$ for $\left<d\right>=6$ than for $\left<d\right>=4$, which suggests that the former might catch up with the latter as the network size increases. On the other hand, the modularity range corresponding to high $\left<p_s\right>$ values seems to become narrower as the average degree increases. These results prove that modularity plays a critical role in a network's ability to support self-sustained oscillation patterns.

The next set of results concerns the relationship between modularity and the average period $\left<P\right>$ of the wave pattern on the network. The period was calculated from the lowest frequency peak (not necessarily the highest) in the discrete Fourier transform of $u_{tot}=\sum_{i=1}^N u_i$ using the last 8192 recorded sets of values. To better correlate the average period with modularity, we considered the union of all network ensembles with given $N$, $\left<d\right>$ and $d_{max}$ but different values of $\mu$ and the resulting range for $Q$ was divided into 10 bins. The period was averaged over all phase space realizations exhibiting self-sustained activity of all networks with modularity within a given bin. It is important to mention that, while the average period varies as shown in Fig.~\ref{fig:Perl-vs-Q}, the individual values of the periods in each modularity bin are distributed over wide intervals, from less than 10 up to hundreds.

Results for networks of size $N=100$ and different coupling strengths $c$ are shown in Fig.~\ref{fig:Perl-vs-Q} (a), while Fig.~\ref{fig:Perl-vs-Q} (b) shows the results for various network sizes $N$ at $c=0.15$. The average degree was $\left<d\right>=4$ while $d_{max}$ was 15 and 16 respectively. Note that the modularity range for $N=800$ in Fig.~\ref{fig:Perl-vs-Q} (b) is narrower than in the case of the other curves. This is because the network generation algorithm breaks down if $\mu\le 0.2$.

While the dependence of the average period on modularity changes in complex ways when the coupling strength is varied, there is a clear trend of overall decrease with increasing $c$, again as a result of the increased susceptibility of the hubs. The curves for $c>0.4$ are statistically indistinguishable from the one for $c=0.4$. The most important feature in Fig.~\ref{fig:Perl-vs-Q} is the presence of a minimum of the average period at medium modularity values, which suggests different mechanisms for the generation of long-period oscillations at the two ends of the modularity range. The minimum is around $Q=0.6$ for $N=100$ but shifts towards higher modularity values as the network size increases.

\begin{figure*}[t]
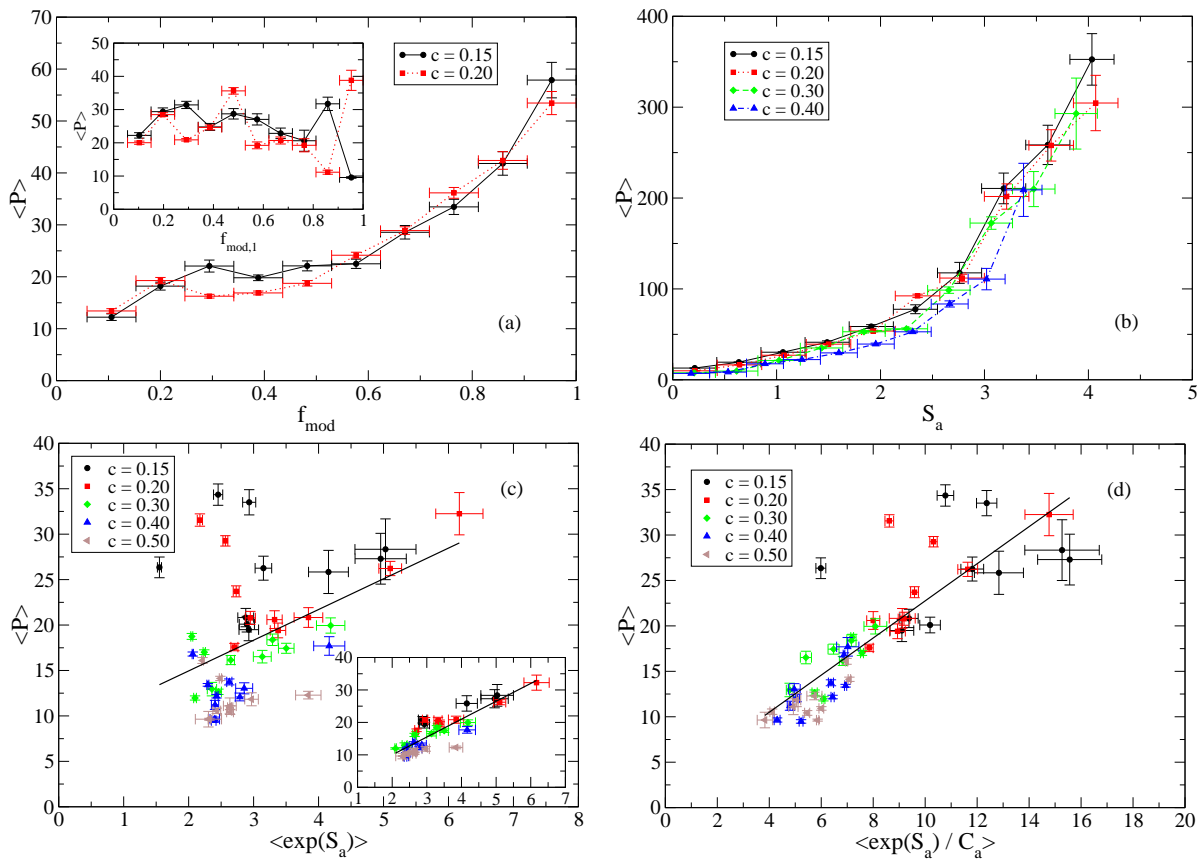

\scalebox{0.32}[0.32]{\includegraphics*{Perl-vs-fml-N100-K400-varc-D2k-s-177290}}
\quad
\scalebox{0.32}[0.32]{\includegraphics*{Perl-vs-Sa-N100-K400-varc-D2k-s-177290}}
\linebreak
\scalebox{0.32}[0.32]{\includegraphics*{Perl-vs-expSa-c-Q-N100-K400-varc-D2k-s-177290alt}}
\quad
\scalebox{0.32}[0.32]{\includegraphics*{Perl-vs-R-c-Q-N100-K400-varc-D2k-s-177290a}}
\caption{\label{fig:Perl-vs-X} (Color online) (a) The average period $\left<P\right>$ for different ranges of the fraction of modules represented in the oscillation source defined using $D_{th}=\min\{D_{i,max}\}$ or DPAD (inset). (b) $\left<P\right>$ for different ranges of the average entropy of the driving spectrum, $S_a$. (c) $\left<P\right>$ versus $\left<exp(S_a)\right>$ for different ranges of modularity. The inset shows the same results but with the high-modularity outliers removed. (d) $\left<P\right>$ versus $\left<exp(S_a)/C_a\right>$ for different ranges of modularity. The vertical bars represent the standard error of the mean while the horizontal bars show the extent of the bins ((a) and (b)) or the standard error of the mean ((c) and (d)). The results are for $N=100$, $\left<d\right>=4$ and different values of the coupling strength $c$.}
\end{figure*}

A first question that arises is whether there are any correlations between period and other quantities characterizing either the topology of the network or the oscillation pattern. Tests failed to reveal any correlation between period and the average degree of the network. Likewise, there is no correlation between period and the size of the oscillation source defined either using individual link thresholds $D_{th,i}=D_{i,max}$ (which is equivalent to the DPAD method) or using a global $D_{th}=\min\{D_{i,max}\}$ (which produces a larger, more detailed subnetwork). However, we found a positive correlation between period and the fraction $f_{mod}$ of modules that have at least one node in the larger oscillation source, as shown in Fig.~\ref{fig:Perl-vs-X} (a), where the average period is plotted for 10 different $f_{mod}$ bins for networks of size $N=100$ and average degree $\left<d\right>=4$. Interestingly, there is no correlation with the fraction of modules $f_{mod,1}$ represented in the smaller, DPAD-defined, source (see inset), which lends additional credibility to the idea that a more detailed description of the source is required. The correlation between period and $f_{mod}$ is present regardless of network size, the value of the coupling strength or, more importantly, the value of modularity and shows that long periods are associated with propagation patterns where most of the modules are involved in the oscillation source.

Inspired by the complexity of the long-period oscillation patterns, which manifests itself among other things in the fact that a given node $i$ does not receive the same total driving $\delta_i=\sum_{j=1}^N \delta_{ij}$ from its neighbors every time $w_i=0$ while increasing, we studied the correlation between period and the entropy of the $\delta_i$ spectrum averaged over all nodes on the network, denoted by $S_a$. Fig.~\ref{fig:Perl-vs-X} (b) shows the average period computed for 10 different bins of $S_a$ and different values of the coupling strength $c$. The correlation between period and average entropy is even stronger than with $f_{mod}$ in the sense that networks with a high average entropy are very likely to exhibit oscillation patterns with a very long period.

A second question is whether we can use this insight to explain the observed dependence of the average period on modularity. To test this, we plotted the average period $\left<P\right>$ versus the average of $exp(S_a)$, both calculated for a given modularity bin and with the error bars denoting the standard error of the mean of either quantity. The exponential of the average entropy can be interpreted as an effective number of distinct $\delta_i$ values. The results, shown in Fig.~\ref{fig:Perl-vs-X} (c) for different values of the coupling strength $c$, do not seem at first to reveal a significant correlation. The Pearson correlation coefficient in this case has a low value of $r=0.438$. However, note that for any value of $c$ the distant outliers are the ones corresponding to the highest modularity bins. By eliminating the points corresponding to the highest 4 bins for $c=0.15$, the highest 3 bins for $c=0.20$ and the highest 2 bins for the other $c$ sets, we find that the correlation improves significantly (see inset), with the Pearson coefficient increasing to $r=0.866$. This amounts to a conclusion that, at least in the case of low- and medium-modularity networks, the period of the oscillatory pattern is reasonably well accounted for by the complexity of the interactions between the nodes on the network. As modularity increases, the complexity of the interactions decreases, which explains the shorter average periods of medium-modularity networks, but fails to explain the long periods of the high-modularity ones.

The explanation for the latter phenomenon seems to reside with the less correlated nature of the oscillations. As modularity increases, the network-wide average $C_a$ of the correlation coefficients between the activities of the various nodes, defined by
\begin{equation}
C_{i,j}=\frac{\left< u_i u_j \right>}{\sqrt{\left< u_i^2 \right> \left< u_j^2 \right>}} ,
\label{eq:four}
\end{equation}
\noindent decreases while the intra-module average increases. Intuitively, a less correlated network will have a longer period because the wave propagates through different parts of the network at different times. A crude attempt to take this into account is to plot $\left<P\right>$ versus the average of $\frac{exp(S_a)}{C_a}$, which is shown in Fig.~\ref{fig:Perl-vs-X} (d). The data points for all modularity bins appear in this figure and, while a few outliers are still present, they are not as far and the overall correlation is significantly improved with $r=0.849$. This result suggests that decreased correlation is indeed responsible for the long average periods of higly modular networks.

\section{Conclusions}

We introduced a new method for identifying the sources of self-sustained oscillations on complex excitable networks, which can be used to analyze them at different levels of detail. Our method is able to provide a complete picture of the interactions resulting in complex oscillation patterns.

We studied the relationship between modularity and the probability of self-sustained oscillations and found that, regardless of coupling strength, high-modularity networks are more likely to be able to support self-sustained oscillations compared to networks of low modularity of the same size and average degree.

We found that both low- and high-modularity networks can support long-period oscillations, but the oscillation patterns in these two cases are qualitatively different, with series of synchronized network-wide bursts exhibiting a multitude of subtle differences in the case of low-modularity networks and series of simpler but more localized bursts in the case of networks with high modularity. The period of the oscillation pattern is statistically correlated with the fraction of modules that are part of the oscillation source, with the entropy of the spectrum of driving values and with the average of the correlation coefficients between the activities of the nodes.

Regardless of modularity, the long-period oscillations cannot be explained by the length of any simple loop on the network, but by interactions between waves propagating along different loops. This proves that the memorization of complex, long-duration patterns does not necessarily require long minimum Winfree loops, as it has been inferred based on Erd\H{o}s-R\'enyi networks and certain simple small-world network models.

Additional research will be required to provide a more precise characterization of the overall oscillation pattern and period of the network in terms of the complexity of the set of propagation times between its parts and the correlations between the activities of the different nodes.

\bibliography{Paper1}

\providecommand{\noopsort}[1]{}\providecommand{\singleletter}[1]{#1}%
\begin{thebibliography}{41}%
\makeatletter
\providecommand \@ifxundefined [1]{%
 \@ifx{#1\undefined}
}%
\providecommand \@ifnum [1]{%
 \ifnum #1\expandafter \@firstoftwo
 \else \expandafter \@secondoftwo
 \fi
}%
\providecommand \@ifx [1]{%
 \ifx #1\expandafter \@firstoftwo
 \else \expandafter \@secondoftwo
 \fi
}%
\providecommand \natexlab [1]{#1}%
\providecommand \enquote  [1]{``#1''}%
\providecommand \bibnamefont  [1]{#1}%
\providecommand \bibfnamefont [1]{#1}%
\providecommand \citenamefont [1]{#1}%
\providecommand \href@noop [0]{\@secondoftwo}%
\providecommand \href [0]{\begingroup \@sanitize@url \@href}%
\providecommand \@href[1]{\@@startlink{#1}\@@href}%
\providecommand \@@href[1]{\endgroup#1\@@endlink}%
\providecommand \@sanitize@url [0]{\catcode `\\12\catcode `\$12\catcode
  `\&12\catcode `\#12\catcode `\^12\catcode `\_12\catcode `\%12\relax}%
\providecommand \@@startlink[1]{}%
\providecommand \@@endlink[0]{}%
\providecommand \url  [0]{\begingroup\@sanitize@url \@url }%
\providecommand \@url [1]{\endgroup\@href {#1}{\urlprefix }}%
\providecommand \urlprefix  [0]{URL }%
\providecommand \Eprint [0]{\href }%
\providecommand \doibase [0]{http://dx.doi.org/}%
\providecommand \selectlanguage [0]{\@gobble}%
\providecommand \bibinfo  [0]{\@secondoftwo}%
\providecommand \bibfield  [0]{\@secondoftwo}%
\providecommand \translation [1]{[#1]}%
\providecommand \BibitemOpen [0]{}%
\providecommand \bibitemStop [0]{}%
\providecommand \bibitemNoStop [0]{.\EOS\space}%
\providecommand \EOS [0]{\spacefactor3000\relax}%
\providecommand \BibitemShut  [1]{\csname bibitem#1\endcsname}%
\let\auto@bib@innerbib\@empty
\bibitem [{\citenamefont {Kapral}(1995)}]{Kapral}%
  \BibitemOpen
  \bibfield  {author} {\bibinfo {author} {\bibfnamefont {R.}~\bibnamefont
  {Kapral}},\ }\href@noop {} {\bibfield  {journal} {\bibinfo  {journal}
  {Physica D}\ }\textbf {\bibinfo {volume} {86}},\ \bibinfo {pages} {149}
  (\bibinfo {year} {1995})}\BibitemShut {NoStop}%
\bibitem [{\citenamefont {Meron}(1992)}]{Meron}%
  \BibitemOpen
  \bibfield  {author} {\bibinfo {author} {\bibfnamefont {E.}~\bibnamefont
  {Meron}},\ }\href@noop {} {\bibfield  {journal} {\bibinfo  {journal} {Physics
  Reports}\ }\textbf {\bibinfo {volume} {218}},\ \bibinfo {pages} {1} (\bibinfo
  {year} {1992})}\BibitemShut {NoStop}%
\bibitem [{\citenamefont {Chen}\ \emph {et~al.}(2014)\citenamefont {Chen},
  \citenamefont {Peng}, \citenamefont {Zhao}, \citenamefont {You},
  \citenamefont {Wu},\ and\ \citenamefont {Ying}}]{ChenEfield}%
  \BibitemOpen
  \bibfield  {author} {\bibinfo {author} {\bibfnamefont {J.}~\bibnamefont
  {Chen}}, \bibinfo {author} {\bibfnamefont {L.}~\bibnamefont {Peng}}, \bibinfo
  {author} {\bibfnamefont {Y.}~\bibnamefont {Zhao}}, \bibinfo {author}
  {\bibfnamefont {S.}~\bibnamefont {You}}, \bibinfo {author} {\bibfnamefont
  {N.}~\bibnamefont {Wu}}, \ and\ \bibinfo {author} {\bibfnamefont
  {H.}~\bibnamefont {Ying}},\ }\href@noop {} {\bibfield  {journal} {\bibinfo
  {journal} {Commun. Nonlin. Sci. Numer. Simulat.}\ }\textbf {\bibinfo {volume}
  {19}},\ \bibinfo {pages} {60} (\bibinfo {year} {2014})}\BibitemShut {NoStop}%
\bibitem [{\citenamefont {B{\"a}r}\ and\ \citenamefont
  {Eiswirth}(1993)}]{BarEiswirth}%
  \BibitemOpen
  \bibfield  {author} {\bibinfo {author} {\bibfnamefont {M.}~\bibnamefont
  {B{\"a}r}}\ and\ \bibinfo {author} {\bibfnamefont {M.}~\bibnamefont
  {Eiswirth}},\ }\href@noop {} {\bibfield  {journal} {\bibinfo  {journal}
  {Phys. Rev. E}\ }\textbf {\bibinfo {volume} {48}},\ \bibinfo {pages} {R1635}
  (\bibinfo {year} {1993})}\BibitemShut {NoStop}%
\bibitem [{\citenamefont {Steinbock}\ \emph {et~al.}(1992)\citenamefont
  {Steinbock}, \citenamefont {Sch{\"u}tze},\ and\ \citenamefont
  {M{\"u}ller}}]{Steinbock}%
  \BibitemOpen
  \bibfield  {author} {\bibinfo {author} {\bibfnamefont {O.}~\bibnamefont
  {Steinbock}}, \bibinfo {author} {\bibfnamefont {J.}~\bibnamefont
  {Sch{\"u}tze}}, \ and\ \bibinfo {author} {\bibfnamefont {S.~C.}\ \bibnamefont
  {M{\"u}ller}},\ }\href@noop {} {\bibfield  {journal} {\bibinfo  {journal}
  {Phys. Rev. Lett.}\ }\textbf {\bibinfo {volume} {68}},\ \bibinfo {pages}
  {248} (\bibinfo {year} {1992})}\BibitemShut {NoStop}%
\bibitem [{\citenamefont {Aranson}\ \emph {et~al.}(1994)\citenamefont
  {Aranson}, \citenamefont {Kramer},\ and\ \citenamefont {Weber}}]{Aranson}%
  \BibitemOpen
  \bibfield  {author} {\bibinfo {author} {\bibfnamefont {I.}~\bibnamefont
  {Aranson}}, \bibinfo {author} {\bibfnamefont {L.}~\bibnamefont {Kramer}}, \
  and\ \bibinfo {author} {\bibfnamefont {A.}~\bibnamefont {Weber}},\
  }\href@noop {} {\bibfield  {journal} {\bibinfo  {journal} {Phys. Rev. Lett.}\
  }\textbf {\bibinfo {volume} {72}},\ \bibinfo {pages} {2316} (\bibinfo {year}
  {1994})}\BibitemShut {NoStop}%
\bibitem [{\citenamefont {Qian}\ and\ \citenamefont
  {Zhang}(2016)}]{QianFundStruct}%
  \BibitemOpen
  \bibfield  {author} {\bibinfo {author} {\bibfnamefont {Y.}~\bibnamefont
  {Qian}}\ and\ \bibinfo {author} {\bibfnamefont {Z.}~\bibnamefont {Zhang}},\
  }\href@noop {} {\bibfield  {journal} {\bibinfo  {journal} {PLoS ONE}\
  }\textbf {\bibinfo {volume} {11}},\ \bibinfo {pages} {e0149842} (\bibinfo
  {year} {2016})}\BibitemShut {NoStop}%
\bibitem [{\citenamefont {Cross}\ and\ \citenamefont
  {Hohenberg}(1993)}]{CrossHohReview}%
  \BibitemOpen
  \bibfield  {author} {\bibinfo {author} {\bibfnamefont {M.~C.}\ \bibnamefont
  {Cross}}\ and\ \bibinfo {author} {\bibfnamefont {P.~C.}\ \bibnamefont
  {Hohenberg}},\ }\href@noop {} {\bibfield  {journal} {\bibinfo  {journal}
  {Rev. Mod. Phys.}\ }\textbf {\bibinfo {volume} {65}},\ \bibinfo {pages} {851}
  (\bibinfo {year} {1993})}\BibitemShut {NoStop}%
\bibitem [{\citenamefont {M{\"u}ller-Linow}\ \emph {et~al.}(2008)\citenamefont
  {M{\"u}ller-Linow}, \citenamefont {Hilgetag},\ and\ \citenamefont
  {H{\"u}tt}}]{MullerStruct}%
  \BibitemOpen
  \bibfield  {author} {\bibinfo {author} {\bibfnamefont {M.}~\bibnamefont
  {M{\"u}ller-Linow}}, \bibinfo {author} {\bibfnamefont {C.~C.}\ \bibnamefont
  {Hilgetag}}, \ and\ \bibinfo {author} {\bibfnamefont {M.-T.}\ \bibnamefont
  {H{\"u}tt}},\ }\href@noop {} {\bibfield  {journal} {\bibinfo  {journal} {PLoS
  Comput. Biol.}\ }\textbf {\bibinfo {volume} {4}},\ \bibinfo {pages}
  {e1000190} (\bibinfo {year} {2008})}\BibitemShut {NoStop}%
\bibitem [{\citenamefont {Kanakov}\ and\ \citenamefont
  {Osipov}(2007)}]{KanakovClustSynch}%
  \BibitemOpen
  \bibfield  {author} {\bibinfo {author} {\bibfnamefont {O.~I.}\ \bibnamefont
  {Kanakov}}\ and\ \bibinfo {author} {\bibfnamefont {G.~V.}\ \bibnamefont
  {Osipov}},\ }\href@noop {} {\bibfield  {journal} {\bibinfo  {journal}
  {Chaos}\ }\textbf {\bibinfo {volume} {17}},\ \bibinfo {pages} {015111}
  (\bibinfo {year} {2007})}\BibitemShut {NoStop}%
\bibitem [{\citenamefont {Ma}\ \emph {et~al.}(2015)\citenamefont {Ma},
  \citenamefont {Song}, \citenamefont {Tang},\ and\ \citenamefont
  {Wang}}]{MaWave}%
  \BibitemOpen
  \bibfield  {author} {\bibinfo {author} {\bibfnamefont {J.}~\bibnamefont
  {Ma}}, \bibinfo {author} {\bibfnamefont {X.}~\bibnamefont {Song}}, \bibinfo
  {author} {\bibfnamefont {J.}~\bibnamefont {Tang}}, \ and\ \bibinfo {author}
  {\bibfnamefont {C.}~\bibnamefont {Wang}},\ }\href@noop {} {\bibfield
  {journal} {\bibinfo  {journal} {Neurocomputing}\ }\textbf {\bibinfo {volume}
  {167}},\ \bibinfo {pages} {378} (\bibinfo {year} {2015})}\BibitemShut
  {NoStop}%
\bibitem [{\citenamefont {Liao}\ \emph {et~al.}(2011)\citenamefont {Liao},
  \citenamefont {Xia}, \citenamefont {Qian}, \citenamefont {Zhang},
  \citenamefont {Hu},\ and\ \citenamefont {Mi}}]{LiaoPatternForm}%
  \BibitemOpen
  \bibfield  {author} {\bibinfo {author} {\bibfnamefont {X.}~\bibnamefont
  {Liao}}, \bibinfo {author} {\bibfnamefont {Q.}~\bibnamefont {Xia}}, \bibinfo
  {author} {\bibfnamefont {Y.}~\bibnamefont {Qian}}, \bibinfo {author}
  {\bibfnamefont {L.}~\bibnamefont {Zhang}}, \bibinfo {author} {\bibfnamefont
  {G.}~\bibnamefont {Hu}}, \ and\ \bibinfo {author} {\bibfnamefont
  {Y.}~\bibnamefont {Mi}},\ }\href@noop {} {\bibfield  {journal} {\bibinfo
  {journal} {Phys. Rev. E}\ }\textbf {\bibinfo {volume} {83}},\ \bibinfo
  {pages} {056204} (\bibinfo {year} {2011})}\BibitemShut {NoStop}%
\bibitem [{\citenamefont {Kakmeni}\ \emph {et~al.}(2014)\citenamefont
  {Kakmeni}, \citenamefont {Inack},\ and\ \citenamefont
  {Yamakou}}]{KakmeniHindmarshRose}%
  \BibitemOpen
  \bibfield  {author} {\bibinfo {author} {\bibfnamefont {F.~M.~M.}\
  \bibnamefont {Kakmeni}}, \bibinfo {author} {\bibfnamefont {E.~M.}\
  \bibnamefont {Inack}}, \ and\ \bibinfo {author} {\bibfnamefont {E.~M.}\
  \bibnamefont {Yamakou}},\ }\href@noop {} {\bibfield  {journal} {\bibinfo
  {journal} {Phys. Rev. E}\ }\textbf {\bibinfo {volume} {89}},\ \bibinfo
  {pages} {052919} (\bibinfo {year} {2014})}\BibitemShut {NoStop}%
\bibitem [{\citenamefont {Zheng}\ and\ \citenamefont {Qian}(2010)}]{LiLiHu}%
  \BibitemOpen
  \bibfield  {author} {\bibinfo {author} {\bibfnamefont {Z.}~\bibnamefont
  {Zheng}}\ and\ \bibinfo {author} {\bibfnamefont {Y.}~\bibnamefont {Qian}},\
  }\href@noop {} {\bibfield  {journal} {\bibinfo  {journal} {Physica A}\
  }\textbf {\bibinfo {volume} {389}},\ \bibinfo {pages} {164} (\bibinfo {year}
  {2010})}\BibitemShut {NoStop}%
\bibitem [{\citenamefont {Riecke}\ \emph {et~al.}(2007)\citenamefont {Riecke},
  \citenamefont {Roxin}, \citenamefont {Madruga},\ and\ \citenamefont
  {Solla}}]{RieckeMultipleAttractors}%
  \BibitemOpen
  \bibfield  {author} {\bibinfo {author} {\bibfnamefont {H.}~\bibnamefont
  {Riecke}}, \bibinfo {author} {\bibfnamefont {A.}~\bibnamefont {Roxin}},
  \bibinfo {author} {\bibfnamefont {S.}~\bibnamefont {Madruga}}, \ and\
  \bibinfo {author} {\bibfnamefont {S.~A.}\ \bibnamefont {Solla}},\ }\href@noop
  {} {\bibfield  {journal} {\bibinfo  {journal} {Chaos}\ }\textbf {\bibinfo
  {volume} {17}},\ \bibinfo {pages} {026110} (\bibinfo {year}
  {2007})}\BibitemShut {NoStop}%
\bibitem [{\citenamefont {Qian}(2015)}]{QianRegulation}%
  \BibitemOpen
  \bibfield  {author} {\bibinfo {author} {\bibfnamefont {Y.}~\bibnamefont
  {Qian}},\ }\href@noop {} {\bibfield  {journal} {\bibinfo  {journal} {Commun
  Nonlinear Sci Numer Simulat}\ }\textbf {\bibinfo {volume} {27}},\ \bibinfo
  {pages} {12} (\bibinfo {year} {2015})}\BibitemShut {NoStop}%
\bibitem [{\citenamefont {Sinha}\ \emph {et~al.}(2007)\citenamefont {Sinha},
  \citenamefont {Saram{\"a}ki},\ and\ \citenamefont
  {Kaski}}]{SinhaEmergenceSW}%
  \BibitemOpen
  \bibfield  {author} {\bibinfo {author} {\bibfnamefont {S.}~\bibnamefont
  {Sinha}}, \bibinfo {author} {\bibfnamefont {J.}~\bibnamefont {Saram{\"a}ki}},
  \ and\ \bibinfo {author} {\bibfnamefont {K.}~\bibnamefont {Kaski}},\
  }\href@noop {} {\bibfield  {journal} {\bibinfo  {journal} {Phys. Rev. E}\
  }\textbf {\bibinfo {volume} {76}},\ \bibinfo {pages} {015101(R)} (\bibinfo
  {year} {2007})}\BibitemShut {NoStop}%
\bibitem [{\citenamefont {Qian}\ \emph {et~al.}(2010)\citenamefont {Qian},
  \citenamefont {Huang}, \citenamefont {Hu},\ and\ \citenamefont
  {Liao}}]{QianStructureControl}%
  \BibitemOpen
  \bibfield  {author} {\bibinfo {author} {\bibfnamefont {Y.}~\bibnamefont
  {Qian}}, \bibinfo {author} {\bibfnamefont {X.}~\bibnamefont {Huang}},
  \bibinfo {author} {\bibfnamefont {G.}~\bibnamefont {Hu}}, \ and\ \bibinfo
  {author} {\bibfnamefont {X.}~\bibnamefont {Liao}},\ }\href@noop {} {\bibfield
   {journal} {\bibinfo  {journal} {Phys. Rev. E}\ }\textbf {\bibinfo {volume}
  {81}},\ \bibinfo {pages} {036101} (\bibinfo {year} {2010})}\BibitemShut
  {NoStop}%
\bibitem [{\citenamefont {Kobayashi}\ \emph {et~al.}(2017)\citenamefont
  {Kobayashi}, \citenamefont {Kitahata},\ and\ \citenamefont
  {Nagayama}}]{KobayashiSustained}%
  \BibitemOpen
  \bibfield  {author} {\bibinfo {author} {\bibfnamefont {Y.}~\bibnamefont
  {Kobayashi}}, \bibinfo {author} {\bibfnamefont {H.}~\bibnamefont {Kitahata}},
  \ and\ \bibinfo {author} {\bibfnamefont {M.}~\bibnamefont {Nagayama}},\
  }\href@noop {} {\bibfield  {journal} {\bibinfo  {journal} {Phys. Rev. E}\
  }\textbf {\bibinfo {volume} {96}},\ \bibinfo {pages} {022213} (\bibinfo
  {year} {2017})}\BibitemShut {NoStop}%
\bibitem [{\citenamefont {Roxin}\ \emph {et~al.}(2004)\citenamefont {Roxin},
  \citenamefont {Riecke},\ and\ \citenamefont {Solla}}]{RoxinSelfSustained}%
  \BibitemOpen
  \bibfield  {author} {\bibinfo {author} {\bibfnamefont {A.}~\bibnamefont
  {Roxin}}, \bibinfo {author} {\bibfnamefont {H.}~\bibnamefont {Riecke}}, \
  and\ \bibinfo {author} {\bibfnamefont {S.~A.}\ \bibnamefont {Solla}},\
  }\href@noop {} {\bibfield  {journal} {\bibinfo  {journal} {Phys. Rev. Lett.}\
  }\textbf {\bibinfo {volume} {92}},\ \bibinfo {pages} {198101} (\bibinfo
  {year} {2004})}\BibitemShut {NoStop}%
\bibitem [{\citenamefont {Qian}(2014)}]{QianER}%
  \BibitemOpen
  \bibfield  {author} {\bibinfo {author} {\bibfnamefont {Y.}~\bibnamefont
  {Qian}},\ }\href@noop {} {\bibfield  {journal} {\bibinfo  {journal} {Phys.
  Rev. E}\ }\textbf {\bibinfo {volume} {90}},\ \bibinfo {pages} {032807}
  (\bibinfo {year} {2014})}\BibitemShut {NoStop}%
\bibitem [{\citenamefont {Qian}\ and\ \citenamefont
  {Zhang}(2017)}]{QianTimeDelay}%
  \BibitemOpen
  \bibfield  {author} {\bibinfo {author} {\bibfnamefont {Y.}~\bibnamefont
  {Qian}}\ and\ \bibinfo {author} {\bibfnamefont {Z.}~\bibnamefont {Zhang}},\
  }\href@noop {} {\bibfield  {journal} {\bibinfo  {journal} {Commun Nonlinear
  Sci Numer Simulat}\ }\textbf {\bibinfo {volume} {47}},\ \bibinfo {pages}
  {127} (\bibinfo {year} {2017})}\BibitemShut {NoStop}%
\bibitem [{\citenamefont {Qian}\ \emph {et~al.}(2017)\citenamefont {Qian},
  \citenamefont {Cui},\ and\ \citenamefont {Zheng}}]{QianMinWinfree}%
  \BibitemOpen
  \bibfield  {author} {\bibinfo {author} {\bibfnamefont {Y.}~\bibnamefont
  {Qian}}, \bibinfo {author} {\bibfnamefont {X.}~\bibnamefont {Cui}}, \ and\
  \bibinfo {author} {\bibfnamefont {Z.}~\bibnamefont {Zheng}},\ }\href@noop {}
  {\bibfield  {journal} {\bibinfo  {journal} {Scientific Reports}\ }\textbf
  {\bibinfo {volume} {7}},\ \bibinfo {pages} {5746} (\bibinfo {year}
  {2017})}\BibitemShut {NoStop}%
\bibitem [{\citenamefont {Avena-Koenigsberger}\ \emph
  {et~al.}(2018)\citenamefont {Avena-Koenigsberger}, \citenamefont {Misic},\
  and\ \citenamefont {Sporns}}]{AvenaCommunDynamics}%
  \BibitemOpen
  \bibfield  {author} {\bibinfo {author} {\bibfnamefont {A.}~\bibnamefont
  {Avena-Koenigsberger}}, \bibinfo {author} {\bibfnamefont {B.}~\bibnamefont
  {Misic}}, \ and\ \bibinfo {author} {\bibfnamefont {O.}~\bibnamefont
  {Sporns}},\ }\href@noop {} {\bibfield  {journal} {\bibinfo  {journal} {Nature
  Reviews Neuroscience}\ }\textbf {\bibinfo {volume} {19}},\ \bibinfo {pages}
  {17} (\bibinfo {year} {2018})}\BibitemShut {NoStop}%
\bibitem [{\citenamefont {Zheng}\ and\ \citenamefont {Qian}(2018)}]{ZhengDPAD}%
  \BibitemOpen
  \bibfield  {author} {\bibinfo {author} {\bibfnamefont {Z.}~\bibnamefont
  {Zheng}}\ and\ \bibinfo {author} {\bibfnamefont {Y.}~\bibnamefont {Qian}},\
  }\href@noop {} {\bibfield  {journal} {\bibinfo  {journal} {Chin. Phys. B}\
  }\textbf {\bibinfo {volume} {27}},\ \bibinfo {pages} {018901} (\bibinfo
  {year} {2018})}\BibitemShut {NoStop}%
\bibitem [{\citenamefont {Bazhenov}\ \emph {et~al.}(1999)\citenamefont
  {Bazhenov}, \citenamefont {Timofeev}, \citenamefont {Steriade},\ and\
  \citenamefont {Sejnowski}}]{BazhenovBrain}%
  \BibitemOpen
  \bibfield  {author} {\bibinfo {author} {\bibfnamefont {M.}~\bibnamefont
  {Bazhenov}}, \bibinfo {author} {\bibfnamefont {I.}~\bibnamefont {Timofeev}},
  \bibinfo {author} {\bibfnamefont {M.}~\bibnamefont {Steriade}}, \ and\
  \bibinfo {author} {\bibfnamefont {T.~J.}\ \bibnamefont {Sejnowski}},\
  }\href@noop {} {\bibfield  {journal} {\bibinfo  {journal} {Nat. Neurosci.}\
  }\textbf {\bibinfo {volume} {2}},\ \bibinfo {pages} {168} (\bibinfo {year}
  {1999})}\BibitemShut {NoStop}%
\bibitem [{\citenamefont {Rulkov}\ \emph {et~al.}(2004)\citenamefont {Rulkov},
  \citenamefont {Timofeev},\ and\ \citenamefont {Bazhenov}}]{RulkovBrain}%
  \BibitemOpen
  \bibfield  {author} {\bibinfo {author} {\bibfnamefont {N.~F.}\ \bibnamefont
  {Rulkov}}, \bibinfo {author} {\bibfnamefont {I.}~\bibnamefont {Timofeev}}, \
  and\ \bibinfo {author} {\bibfnamefont {M.}~\bibnamefont {Bazhenov}},\
  }\href@noop {} {\bibfield  {journal} {\bibinfo  {journal} {J. Comput.
  Neurosci.}\ }\textbf {\bibinfo {volume} {17}},\ \bibinfo {pages} {203}
  (\bibinfo {year} {2004})}\BibitemShut {NoStop}%
\bibitem [{\citenamefont {Buzs\'{a}ki}\ and\ \citenamefont
  {Draguhn}(2004)}]{BuzsakiBrain}%
  \BibitemOpen
  \bibfield  {author} {\bibinfo {author} {\bibfnamefont {G.}~\bibnamefont
  {Buzs\'{a}ki}}\ and\ \bibinfo {author} {\bibfnamefont {A.}~\bibnamefont
  {Draguhn}},\ }\href@noop {} {\bibfield  {journal} {\bibinfo  {journal}
  {Science}\ }\textbf {\bibinfo {volume} {304}},\ \bibinfo {pages} {1926}
  (\bibinfo {year} {2004})}\BibitemShut {NoStop}%
\bibitem [{\citenamefont {Usrey}\ and\ \citenamefont
  {Reid}(1999)}]{UsreyVisual}%
  \BibitemOpen
  \bibfield  {author} {\bibinfo {author} {\bibfnamefont {W.~M.}\ \bibnamefont
  {Usrey}}\ and\ \bibinfo {author} {\bibfnamefont {R.~C.}\ \bibnamefont
  {Reid}},\ }\href@noop {} {\bibfield  {journal} {\bibinfo  {journal} {Annu.
  Rev. Physiol.}\ }\textbf {\bibinfo {volume} {61}},\ \bibinfo {pages} {435}
  (\bibinfo {year} {1999})}\BibitemShut {NoStop}%
\bibitem [{\citenamefont {Ward}(2003)}]{WardCognitive}%
  \BibitemOpen
  \bibfield  {author} {\bibinfo {author} {\bibfnamefont {L.~M.}\ \bibnamefont
  {Ward}},\ }\href@noop {} {\bibfield  {journal} {\bibinfo  {journal} {Trends
  Cognit. Sci.}\ }\textbf {\bibinfo {volume} {7}},\ \bibinfo {pages} {553}
  (\bibinfo {year} {2003})}\BibitemShut {NoStop}%
\bibitem [{\citenamefont {Steriade}\ \emph {et~al.}(1993)\citenamefont
  {Steriade}, \citenamefont {McCormick},\ and\ \citenamefont
  {Sejnowski}}]{SteriadeSleepArousal}%
  \BibitemOpen
  \bibfield  {author} {\bibinfo {author} {\bibfnamefont {M.}~\bibnamefont
  {Steriade}}, \bibinfo {author} {\bibfnamefont {D.~A.}\ \bibnamefont
  {McCormick}}, \ and\ \bibinfo {author} {\bibfnamefont {T.~J.}\ \bibnamefont
  {Sejnowski}},\ }\href@noop {} {\bibfield  {journal} {\bibinfo  {journal}
  {Science}\ }\textbf {\bibinfo {volume} {262}},\ \bibinfo {pages} {679}
  (\bibinfo {year} {1993})}\BibitemShut {NoStop}%
\bibitem [{\citenamefont {Timme}\ \emph {et~al.}(2016)\citenamefont {Timme},
  \citenamefont {Ito}, \citenamefont {Myroshnychenko}, \citenamefont {Nigam},
  \citenamefont {Shimono}, \citenamefont {Yeh}, \citenamefont {Hottowy},
  \citenamefont {Litke},\ and\ \citenamefont {Beggs}}]{TimmeCorticalComput}%
  \BibitemOpen
  \bibfield  {author} {\bibinfo {author} {\bibfnamefont {N.~M.}\ \bibnamefont
  {Timme}}, \bibinfo {author} {\bibfnamefont {S.}~\bibnamefont {Ito}}, \bibinfo
  {author} {\bibfnamefont {M.}~\bibnamefont {Myroshnychenko}}, \bibinfo
  {author} {\bibfnamefont {S.}~\bibnamefont {Nigam}}, \bibinfo {author}
  {\bibfnamefont {M.}~\bibnamefont {Shimono}}, \bibinfo {author} {\bibfnamefont
  {F.-C.}\ \bibnamefont {Yeh}}, \bibinfo {author} {\bibfnamefont
  {P.}~\bibnamefont {Hottowy}}, \bibinfo {author} {\bibfnamefont {A.~M.}\
  \bibnamefont {Litke}}, \ and\ \bibinfo {author} {\bibfnamefont {J.~M.}\
  \bibnamefont {Beggs}},\ }\href@noop {} {\bibfield  {journal} {\bibinfo
  {journal} {PLoS Comput Biol}\ }\textbf {\bibinfo {volume} {12}},\ \bibinfo
  {pages} {e1004858} (\bibinfo {year} {2016})}\BibitemShut {NoStop}%
\bibitem [{\citenamefont {Mi}\ \emph {et~al.}(2013)\citenamefont {Mi},
  \citenamefont {Liao}, \citenamefont {Huang}, \citenamefont {Zhang},
  \citenamefont {Gu}, \citenamefont {Hu},\ and\ \citenamefont
  {Wu}}]{MiLongPeriodPNAS}%
  \BibitemOpen
  \bibfield  {author} {\bibinfo {author} {\bibfnamefont {Y.}~\bibnamefont
  {Mi}}, \bibinfo {author} {\bibfnamefont {X.}~\bibnamefont {Liao}}, \bibinfo
  {author} {\bibfnamefont {X.}~\bibnamefont {Huang}}, \bibinfo {author}
  {\bibfnamefont {L.}~\bibnamefont {Zhang}}, \bibinfo {author} {\bibfnamefont
  {W.}~\bibnamefont {Gu}}, \bibinfo {author} {\bibfnamefont {G.}~\bibnamefont
  {Hu}}, \ and\ \bibinfo {author} {\bibfnamefont {S.}~\bibnamefont {Wu}},\
  }\href@noop {} {\bibfield  {journal} {\bibinfo  {journal} {PNAS}\ }\textbf
  {\bibinfo {volume} {110}},\ \bibinfo {pages} {E4931} (\bibinfo {year}
  {2013})}\BibitemShut {NoStop}%
\bibitem [{\citenamefont {Lancichinetti}\ \emph {et~al.}(2008)\citenamefont
  {Lancichinetti}, \citenamefont {Fortunato},\ and\ \citenamefont
  {Radicchi}}]{Benchmarks}%
  \BibitemOpen
  \bibfield  {author} {\bibinfo {author} {\bibfnamefont {A.}~\bibnamefont
  {Lancichinetti}}, \bibinfo {author} {\bibfnamefont {S.}~\bibnamefont
  {Fortunato}}, \ and\ \bibinfo {author} {\bibfnamefont {F.}~\bibnamefont
  {Radicchi}},\ }\href@noop {} {\bibfield  {journal} {\bibinfo  {journal}
  {Phys. Rev. E}\ }\textbf {\bibinfo {volume} {78}},\ \bibinfo {pages} {046110}
  (\bibinfo {year} {2008})}\BibitemShut {NoStop}%
\bibitem [{\citenamefont {Fortunato}(2010)}]{FortuRev}%
  \BibitemOpen
  \bibfield  {author} {\bibinfo {author} {\bibfnamefont {S.}~\bibnamefont
  {Fortunato}},\ }\href@noop {} {\bibfield  {journal} {\bibinfo  {journal}
  {Physics Reports}\ }\textbf {\bibinfo {volume} {486}},\ \bibinfo {pages} {75}
  (\bibinfo {year} {2010})}\BibitemShut {NoStop}%
\bibitem [{\citenamefont {Newman}\ and\ \citenamefont
  {Girvan}(2004)}]{NewmanDivisive}%
  \BibitemOpen
  \bibfield  {author} {\bibinfo {author} {\bibfnamefont {M.~E.~J.}\
  \bibnamefont {Newman}}\ and\ \bibinfo {author} {\bibfnamefont
  {M.}~\bibnamefont {Girvan}},\ }\href@noop {} {\bibfield  {journal} {\bibinfo
  {journal} {Phys. Rev. E}\ }\textbf {\bibinfo {volume} {69}},\ \bibinfo
  {pages} {026113} (\bibinfo {year} {2004})}\BibitemShut {NoStop}%
\bibitem [{\citenamefont {Newman}(2006{\natexlab{a}})}]{NewmanPREspect}%
  \BibitemOpen
  \bibfield  {author} {\bibinfo {author} {\bibfnamefont {M.~E.~J.}\
  \bibnamefont {Newman}},\ }\href@noop {} {\bibfield  {journal} {\bibinfo
  {journal} {Phys. Rev. E}\ }\textbf {\bibinfo {volume} {74}},\ \bibinfo
  {pages} {036104} (\bibinfo {year} {2006}{\natexlab{a}})}\BibitemShut
  {NoStop}%
\bibitem [{\citenamefont {Newman}(2006{\natexlab{b}})}]{NewmanPNAS}%
  \BibitemOpen
  \bibfield  {author} {\bibinfo {author} {\bibfnamefont {M.~E.~J.}\
  \bibnamefont {Newman}},\ }\href@noop {} {\bibfield  {journal} {\bibinfo
  {journal} {Proc. Natl. Acad. Sci. USA}\ }\textbf {\bibinfo {volume} {103}},\
  \bibinfo {pages} {8577} (\bibinfo {year} {2006}{\natexlab{b}})}\BibitemShut
  {NoStop}%
\bibitem [{\citenamefont {Newman}(2004)}]{NewmanGreedy}%
  \BibitemOpen
  \bibfield  {author} {\bibinfo {author} {\bibfnamefont {M.~E.~J.}\
  \bibnamefont {Newman}},\ }\href@noop {} {\bibfield  {journal} {\bibinfo
  {journal} {Phys. Rev. E}\ }\textbf {\bibinfo {volume} {69}},\ \bibinfo
  {pages} {066133} (\bibinfo {year} {2004})}\BibitemShut {NoStop}%
\bibitem [{\citenamefont {Jahnke}\ and\ \citenamefont
  {Winfree}(1991)}]{WinfreeSurveyBehav}%
  \BibitemOpen
  \bibfield  {author} {\bibinfo {author} {\bibfnamefont {W.}~\bibnamefont
  {Jahnke}}\ and\ \bibinfo {author} {\bibfnamefont {A.~T.}\ \bibnamefont
  {Winfree}},\ }\href@noop {} {\bibfield  {journal} {\bibinfo  {journal} {Int.
  J. Bifurcation Chaos Appl. Sci. Eng.}\ }\textbf {\bibinfo {volume} {01}},\
  \bibinfo {pages} {445} (\bibinfo {year} {1991})}\BibitemShut {NoStop}%
\bibitem [{Vid()}]{Video}%
  \BibitemOpen
  \href@noop {} {}\bibinfo {note} {See the Phys. Rev. E article for a video of
  the simulation. Note that the counter labeled ``Time'' actually counts the
  frames.}\BibitemShut {Stop}%
\end{thebibliography}%

\end{document}